\documentclass[aps,prl,twocolumn,superscriptaddress]{revtex4}

\usepackage{amsmath}
\usepackage{graphicx}
\usepackage{dcolumn}
\usepackage{sidecap}
\newcommand{\mypath}[1]{./#1}

\begin{document}

\title{Modeling of Fe pnictides: the Magnetic Order and Pairing Channels}

\author{M. Daghofer}
\affiliation{Department of Physics and Astronomy, The University of Tennessee, Knoxville, TN 37996}
\affiliation{Materials Science and Technology Division, Oak Ridge National Laboratory, Oak Ridge, TN 32831}
 
\author{A. Moreo}
\affiliation{Department of Physics and Astronomy, The University of Tennessee, Knoxville, TN 37996}
\affiliation{Materials Science and Technology Division, Oak Ridge National Laboratory, Oak Ridge, TN 32831}

\author{J. A. Riera}
\affiliation{Instituto de F\'isica Rosario, Consejo Nacional de Investigaciones Cient\'ificas y T\'ecnicas, Universidad Nacional de Rosario, 2000-Rosario, Argentina}

\author{E. Arrigoni}
\affiliation{Institute of Theoretical and Computational Physics, TU Graz, A-8010 Graz, Austria}

\author{D.J. Scalapino}
\affiliation{Department of Physics, University of California, Santa Barbara, CA 93106-9530}

\author{E. Dagotto}
\affiliation{Department of Physics and Astronomy, The University of Tennessee, Knoxville, TN 37996}
\affiliation{Materials Science and Technology Division, Oak Ridge National Laboratory, Oak Ridge, TN 32831}

\date{\today}

\begin{abstract}
A two-orbital model for Fe-pnictide superconductors 
is investigated using computational techniques on two-dimensional square clusters.
The hopping amplitudes are derived from orbital overlap integrals, or by band structure fits,
and the spin frustrating effect of the plaquette-diagonal Fe-Fe hopping 
is remarked. A spin ``striped'' state is 
stable in a broad range of couplings in the undoped
regime, in agreement with neutron scattering. 
Adding two electrons to the undoped ground state of a small cluster,
the dominant pairing operators are found.
Depending on parameters, two pairing operators were identified:
they involve inter-$xz$-$yz$ orbital combinations forming spin singlets or triplets, 
transforming according to the $B_{\rm 2g}$ and $A_{\rm 2g}$ representations of the $D_{\rm 4h}$ group, respectively.
\end{abstract}


\maketitle

{\it Introduction:}
The recent discovery of superconductivity  in the layered rare-earth oxypnictides compounds
$\rm LnO_{\it 1-x}F_{\it x}FeAs$ (Ln=La, Pr, Ce, Sm) has captured the attention
of the condensed matter community \cite{Fe-SC}. 
The high current record critical temperature $T_c\sim$55~K  in $\rm SmO_{\it 1-x}F_{\it x}FeAs$ \cite{55} suggests
that an unconventional pairing mechanism may be at work \cite{sefat,boeri}.

As for Cu-based high temperature superconductors (HTSC), the analysis
of undoped compounds, such as $\rm LaOFeAs$, 
is expected to provide important information
toward the understanding of the superconducting (SC) state reached by $\sim$10\% F doping.
Neutron scattering experiments have provided evidence of
magnetic order in $\rm LaOFeAs$ at 134~K: Fe spins
order into ferromagnetic ``stripes'' that are aligned
antiferromagnetically \cite{dai,sdw,others}. 
In the two-dimensional (2D) square lattice notation, the $\rm LaOFeAs$ 
magnetic structure factor has
peaks at wavevectors $q$$\sim$$(0,\pi)$,$(\pi,0)$ \cite{dai,sdw,others}. 
Assuming a smooth continuity
between the undoped and F-doped compounds, the pairing mechanism could be
magnetic in origin and triggered by this unusual magnetic state.

Theoretical work on the new superconductors includes 
band structure calculations that have shown the relevance of the $3d$ levels of Fe~\cite{lda,cao}. 
A metallic state involving 
a  Fermi surface (FS) made out of disconnected
small pieces (``pockets'') was predicted \cite{lda}.
To understand some of the properties 
of the undoped limit, electron correlations 
appear to be important \cite{haule}. 
Two-orbitals descriptions \cite{scalapino,li} and other models
have been proposed, and a variety of approximations have
lead to several
unconventional pairing channel proposals~\cite{kuroki,extra,FCZhang}.

Our purpose is to report the first unbiased computational results obtained 
using a model Hamiltonian
for Fe pnictides, with emphasis on a real-space description.  
In the undoped limit, a spin striped magnetic state 
is obtained and explained. With light electron doping, novel pairing
operators are identified. 
The path followed here mimics research in the HTSC,
where the computational study of model Hamiltonians in real space \cite{RMP} provided a
perspective dual to 
momentum-space diagrammatic calculations. In fact, early numerical studies of the 2-hole
state on small $t$-$J$ clusters indicated that the pairing was in the 
$d_{\rm x^2-y^2}$ channel \cite{RMP}. Thus, it is natural to follow 
a similar path for the new Fe superconductors. 

\begin{figure}
\centerline{
\includegraphics[clip, height=6.0cm]{\mypath{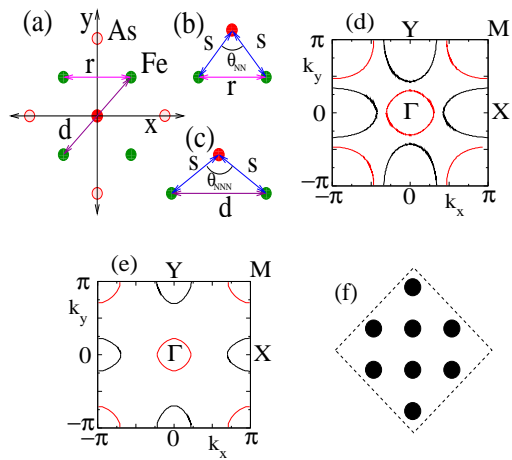}}}
\caption{(Color online)
(a) Small cluster illustrating the geometry of the FeAs layer. 
Open (filled) green circles indicate As positions
above (below) the Fe plane. (b) The Fe-Fe NN path. (c) The Fe-Fe NNN path. 
(d) Fermi surface of $H_{\rm FeAs}$ 
for $pd\pi/pd\sigma$=$-0.2$ in
the $U=J=0$ limit. 
(e) Same as (d) but for the parameters of Ref.~\cite{scalapino}. 
In (d), the half-filled chemical potential is at $-0.03$.
(f) The tilted 8-site cluster used here.
}
\label{Figure1}
\end{figure}

{\it Model and Techniques:} 
Fe pnictides have a layered structure, with the Fe 
atoms forming a 2D square lattice and the As atoms located 
above or below the plane, at the centers of alternating plaquettes, see Fig.~1(a). 
Here, the emphasis
will be on the Fe $d_{\rm xz}$ and $d_{\rm yz}$ degenerate states \cite{scalapino,li}
since band
structure calculations have shown the importance of these orbitals at the
Fermi surface \cite{boeri,bands}. Since the complexity of the problem rapidly increases
with the number of orbitals, it is reasonable 
to start with just two orbitals, contrast the results
with experiments, and slowly build up a more realistic model with extra orbitals.
To estimate the hopping amplitudes,
the Slater-Koster (SK) tight-binding scheme is followed \cite{slater}. 
This approach is simple, analytical, 
and
leads to
a geometrical understanding of the magnetic phase.
The SK method for the hopping integrals needs as input the  
location of the Fe and As atoms, and the nature of the orbitals. The Fe-Fe (Fe-As) distance
used is $r$=2.854 $\rm \AA$ ($s$=2.327 $\rm \AA$). 
The effective Fe-Fe hopping amplitudes - via As - are 
the product of the direct Fe-As hoppings, the three $p$ orbitals were taken
into account on the intermediate As ion. 
Two Fe-As-Fe paths connect nearest-neighbor (NN) Fe-sites, while only
one exists for next-nearest-neighbor (NNN) Fe's along the plaquette
diagonals. 
The kinetic energy 
of the resulting model restricted 
to Fe sites is
\begin{eqnarray}\begin{split}\label{kin}
H_{\rm k} & = -t_1\sum_{{\bf i},\sigma}
(d^\dagger_{{\bf i},x,\sigma}d^{\phantom{\dagger}}_{{\bf i}+{\hat x},x,\sigma} +
 d^\dagger_{{\bf i},y,\sigma}d^{\phantom{\dagger}}_{{\bf i}+{\hat y},y,\sigma} +
\textrm{H. c.})\\
&\quad  -t_2\sum_{{\bf i},\sigma}
(d^\dagger_{{\bf i},y,\sigma}d^{\phantom{\dagger}}_{{\bf i}+{\hat x},y,\sigma} +
 d^\dagger_{{\bf i},x,\sigma}d^{\phantom{\dagger}}_{{\bf i}+{\hat y},x,\sigma} +
\textrm{H. c.})\\
&\quad - t_3\sum_{{\bf i},\sigma}
( d^\dagger_{{\bf i},x,\sigma}d^{\phantom{\dagger}}_{{\bf i}+{\hat x}+{\hat y},x,\sigma} 
+ d^\dagger_{{\bf i},x,\sigma}d^{\phantom{\dagger}}_{{\bf i}+{\hat x}-{\hat y},x,\sigma}\\ 
& \quad\quad+ d^\dagger_{{\bf i},y,\sigma}d^{\phantom{\dagger}}_{{\bf i}+{\hat x}+{\hat y},y,\sigma} 
+ d^\dagger_{{\bf i},y,\sigma}d^{\phantom{\dagger}}_{{\bf i}+{\hat x}-{\hat y},y,\sigma} 
+ \textrm{H. c.})\\
&\quad - t_4\sum_{{\bf i},\sigma}
( d^\dagger_{{\bf i},x,\sigma}d^{\phantom{\dagger}}_{{\bf i}+{\hat x}+{\hat y},y,\sigma} 
+ d^\dagger_{{\bf i},y,\sigma}d^{\phantom{\dagger}}_{{\bf i}+{\hat x}+{\hat y},x,\sigma} 
+ \textrm{H. c.})\\
&\quad + t_4\sum_{{\bf i},\sigma}
( d^\dagger_{{\bf i},x,\sigma}d^{\phantom{\dagger}}_{{\bf i}+{\hat x}-{\hat y},y,\sigma} 
+ d^\dagger_{{\bf i},y,\sigma}d^{\phantom{\dagger}}_{{\bf i}+{\hat x}-{\hat y},x,\sigma} 
+ \textrm{H. c.}),
\end{split}
\end{eqnarray}
\noindent where $d^\dagger_{{\bf i},\alpha,\sigma}$ creates an
electron with spin $\sigma$ in the orbitals $\alpha$=$x$,$y$ 
($d_{\rm xz}$ and  $d_{\rm yz}$, respectively) at site ${\bf i}$ of
a 2D square lattice. $\hat x$ and $\hat y$ are unit vectors along the axes.
The SK-evaluated Fe-Fe hopping amplitudes are 
$t_1$=$-2[(b^2-a^2)+g^2]/ \Delta_{pd}$,
$t_2$=$-2[(b^2-a^2)-g^2]/ \Delta_{pd}$,
$t_3$=$-(a^2+b^2-g^2)/ \Delta_{pd}$, and
 $t_4$=$-(ab-g^2)/ \Delta_{pd}$, where the Fe-As hopping amplitudes are
$a$=$0.324(pd\sigma)-0.374(pd\pi)$,
$b$=$0.324(pd\sigma)+0.123(pd\pi)$, and
$g$=$0.263(pd\sigma)+0.31(pd\pi)$. $pd\sigma$ and $pd\pi$ are SK
parameters and $\Delta_{pd}$ is the energy difference between the $p$
and $d$ levels. The overall energy scale is set by
$(pd\sigma)^2/\Delta_{pd}$, which is of the order of eV and will be
used as unit of energy.
$pd\pi/pd\sigma$ is a free parameter in $H_{\rm k}$. Equation~(1) is
formally the same as in Refs.~\cite{scalapino,li}, but the values for 
the hoppings are different: our approach relies on the
analytic calculation of the hoppings in a ``first-principles'' SK-based context,
while Ref.~\cite{scalapino} fits the hoppings to bands from LDA calculations.
An interesting conclusion of our effort is that both sets of parameters lead
to similar results for the magnetic order and the pairing. 
Equation (1) has invariance under the $D_{\rm 4h}$ point-group
\cite{wan}, including  a 
$\pi/2$  rotation of the lattice together with  
orbital exchanges $x \rightarrow y$ and $y\rightarrow -x$.

The on-site Coulombic terms include a Hubbard
repulsion $U$ for electrons with the same $\alpha$, 
a repulsion $U'$ for different $\alpha$, 
a ferromagnetic Hund coupling $J$, and a pair-hopping 
term with strength $J'$=$J$ \cite{oles}:
\begin{eqnarray}\begin{split}\label{eq:int}
& H_\textrm{int} = U\sum_{{\bf i},\alpha}
n_{{\bf i},\alpha,\uparrow}n_{{\bf i},\alpha,\downarrow} + (U'-J/2)\sum_{{\bf i}}n_{{\bf i},x}n_{{\bf i},y}  \\
&- 2J \sum_{{\bf i}}{{\bf S}_{{\bf i},x}}\cdot{{\bf S}_{{\bf i},y}}
+ J\sum_{{\bf i}}
(d^\dagger_{{\bf i},x,\uparrow}d^\dagger_{{\bf i},x,\downarrow}
d^{\phantom{\dagger}}_{{\bf i},y,\downarrow}
d^{\phantom{\dagger}}_{{\bf i},y,\uparrow}
+ \textrm{H. c.}).
\end{split}\end{eqnarray}
${\bf S}_{{\bf i},\alpha}$ ($n_{{\bf i},\alpha}$) 
is the spin (density) in orbital $\alpha$ at site
${\bf i}$. The standard relation $U'$=$U -2J$ due to rotational invariance 
was used \cite{hotta}. 
The full model becomes $H_{\rm FeAs}$=$H_{\rm k }$+$H_{\rm int}$.
Since the NN Fe-As-Fe bond angle $\theta_{\rm NN}$ [Fig.~1(b)] 
is closer to $90^\circ$ than the NNN Fe-As-Fe angle $\theta_{\rm NNN}$
[Fig.~1(c)], we find a strong NNN hopping $t_3$ and the ratio $t_3/t_1$ is
of order 1 for broad ranges of $pd\pi/pd\sigma$, without fine tuning. 
At intermediate to large $U$, the resulting effective Fe-Fe spin interaction
along the plaquette diagonals consequently becomes as large 
as between NN Fe sites \cite{extra}, or even larger. 
In the early days of HTSC, investigations of the resulting frustrated
effective spin model
unveiled a spin striped phase in the one-orbital model
\cite{adriana}. 
As shown in Figs.~1(d,e), the non-interacting system has
electron FS around the $(X,Y)$ points, and hole FS around the $\Gamma$
and $M$ points, which are equivalent on folding the Brillouin zone. 
2-orbital models cannot have both hole pockets around $\Gamma$ point,
as found in band structure calculations~\cite{lda,cao,scalapino}.



{\it Magnetic properties in the undoped limit:}
We study the ground state of model Eqs.(1,2) in the undoped limit by using 
two techniques: Exact Diagonalization (ED) and the
Variational Cluster Approach (VCA). The first method allows for an unbiased
analysis, albeit restricted to small clusters \cite{RMP}, 
while the second extends the calculation
to the bulk self-consistently \cite{Aic03,Pot03}. We apply ED to the
2$\times$2 and tilted $\sqrt{8}\times\sqrt{8}$ [Fig.~1(f)] 
clusters with periodic boundary conditions~\cite{RMP}. 
While the 2$\times$2 cluster only has 4,900
states even if no symmetries are used, the 8-sites cluster has 
20,706,468 states with translational invariance implemented, and is 
computationally demanding. We therefore fixed the ratio $U/J$ to 4, compatible
with some estimates \cite{cao}. 
However,  other
$U/J$ rations (to be discussed in future publications) do not
critically affect the results presented below. 
In particular, 
the spin striped state and the singlet pairing 
(discussed later) survive for small $J$.  
The typical inequality $|pd\pi/pd\sigma|< 1$ is assumed, and the sign of $pd\pi$ is
chosen such that the FS agrees with band structure calculations (see below).
Both on the 2$\times$2 and on the $\sqrt{8}\times\sqrt{8}$, we observe 
magnetic order with $q=(0,\pi),(\pi,0)$ in 
real-space spin correlations as well as in the magnetic structure
factor $S(q)$, see, e.g., Fig.~2(a). This leads us to believe that size effects are not severe. 
The
$q$=$(0,\pi),(\pi,0)$ state is stable at least in the large square
$-0.5<pd\pi/pd\sigma<0$ and $0<U<4$, and it is  
generated by the robust plaquette-diagonal hoppings. 
We find similar results for
both the SK hoppings and those of Ref.~\cite{scalapino}, and our
results also agree with weak-coupling RPA approximations, where
similar order arises from  nesting \cite{scalapino,yildirim}.
\begin{figure}
\centerline{
\includegraphics[clip, height=7.0cm]{\mypath{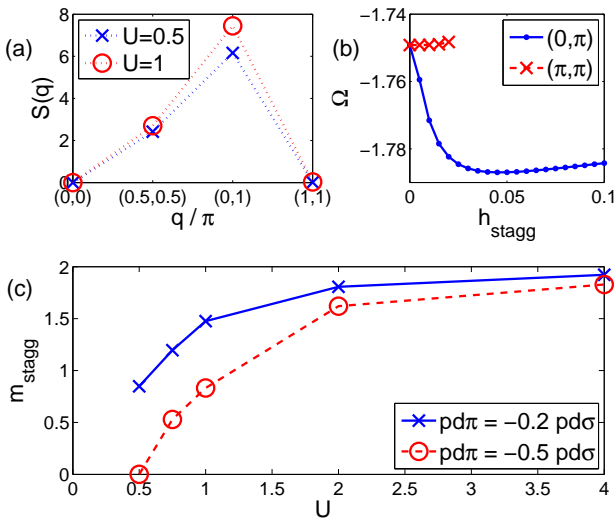}}}
\caption{(Color online) 
(a) $S(q)$ vs. $q$ for the $\sqrt{8}\times\sqrt{8}$ 
cluster at the $U$'s indicated, with $J=U/4$ and $pd\pi/pd\sigma$=$-0.2$. The momenta allowed in this
cluster are $(0,0)$, $(\pm \pi/2, \pm \pi/2)$, $(0,\pi)$, $(\pi,0)$, and $(\pi,\pi)$ \cite{dago92}.
(b) VCA grand potential 
[in energy units $(pd\sigma)^2/\Delta_{pd}$] 
vs. staggered magnetic fields $h_{\rm stagg}$ 
for ${\bf q} = (\pi,\pi)$ and $(0,\pi)$, 
and $pd\pi/pd\sigma$=$-0.2$, $U=1$, $J=0.25$,
The minimum for $(0,\pi)$ at $h_{\rm stagg}\neq0$ indicates symmetry breaking.
(c) $m_{\rm stagg}$ vs. $U$, with $U/J=4$. The small $U$ region at $pd\pi/pd\sigma$=$-0.2$ was numerically unstable.
}
\label{Figure2}
\end{figure}
Further evidence that we indeed identified the dominant magnetic
channel comes from the VCA~\cite{Aic03}, where the self-energy of a small cluster is
optimized by varying appropriate ``fictitious'' fields such as chemical potentials
or symmetry-breaking staggered magnetic fields \cite{Aic03,Pot03}.
It thus combines the exact solution of a small cluster with access to the bulk limit \cite{VCAcomment}.
The grand potential [Fig.~2(b)] demonstrates that the symmetry breaking indeed
occurs in the $q$=$(0,\pi),(\pi,0)$ channel.
Figure~2(c) shows the stripe order-parameter $m_{\rm stagg}$ (the
staggered moment in units of Bohr magneton per Fe ion) vs. $U$: we find
a ($U$,~$pd\pi/pd\sigma$) regime that can accommodate the small $m_{\rm stagg}$ 
found with  neutrons \cite{dai}. Note also that other experiments
have reported larger $m_{\rm stagg}$ values \cite{others}. 


Figure~3 shows photoemission spectra 
$A({\bf k},\omega)$. 
The first case (a) for $pd\pi/pd\sigma=-0.2, U=0.5$ has a dispersion similar to that of
the non-interacting system [Fig.~1(d)], its FS is shown in
Fig.~3(d) and the nodal structure will be 
discussed in a future publication. The density of states (DOS) has a small 
pseudogap [Fig.~3(c)], somewhat deeper than for $U=0$, while larger
$U=2$ leads to an insulating hard gap.
Another interesting regimes is shown in (b): At $U=1$ and $pd\pi/pd\sigma$=$-0.5$,
the chemical potential lies in a region with many states, suggesting a
correlated metal.
%
\begin{figure}
\centerline{
\includegraphics[clip,height=7.5cm]{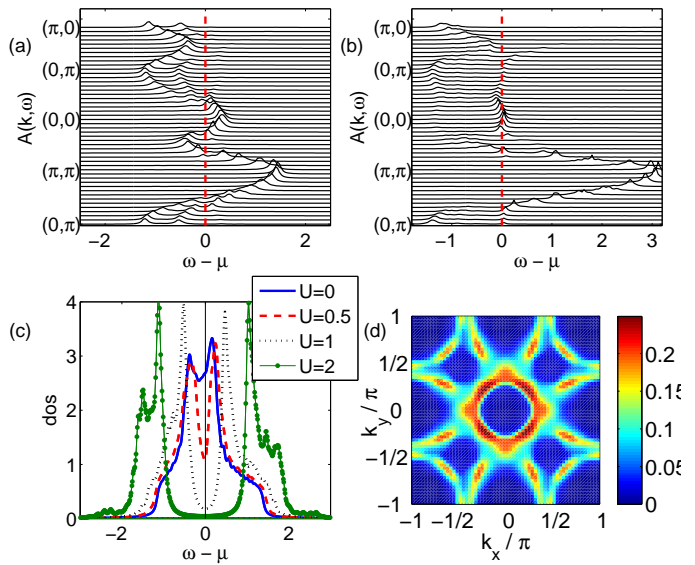}}
\caption{(Color online) 
One-particle VCA spectral function $A({\bf k},\omega)$ 
of $H_{\rm FeAs}$ for (a) $U$=$0.5$, $J$=$0.125$, and $pd\pi/pd\sigma$=$-0.2$,
and (b) $U$=$1$, $J$=$0.25$, $pd\pi/pd\sigma$=$-0.5$ in the symmetry broken phase
with $(0,\pi)$ magnetic ordering at half-filling. A broadening 0.05
was used.  
(c) Density-of-states obtained by ${\bf k}$-integrating 
the spectral functions, at the $U$'s indicated, with $U/J$=$4$ 
and $pd\pi/pd\sigma$=$-0.2$. 
(d) FS corresponding to case (a). The FS has
been symmetrized under rotations. 
}
\label{Figure3}
\end{figure}

{\it Pairing channels:} 
In the HTSC cuprates, the dominant pairing channel could be identified 
by adding two carriers to the undoped finite cluster (which has the same
symmetries as the bulk system) and by then 
evaluating the quantum numbers under $\pi/2$ rotations~\cite{RMP}. 
Since NNN hoppings play a key role in the present case, we need at least a  2$\times$2 cluster
for \emph{each sublattice}, the minimal cluster
satisfying the requirement is the  $\sqrt{8}\times\sqrt{8}$ cluster. 
Varying $U$ and $pd\pi/pd\sigma$ ($U/J$=$4$), we find that 
both singlet and triplet regimes can be reached by adding two electrons to the
undoped (i.e. half-filled) system~\cite{bound}, see the 
phase diagram in Fig.~4(a). At small $|pd\pi/pd\sigma|$ and
intermediate or large $U$, the total spin is 0, and 
in searching for the local operator connecting the ground states of the
undoped and doped systems, the largest overlap at intermediate $U$ is for  
\begin{eqnarray}\begin{split}\label{eq:pair_singlet}
&\rm\Delta^{\dagger}({\bf i})=\sum_{\alpha,{\hat \mu}}
(d^{\dagger}_{{\bf{i}},\alpha,\uparrow} d^{\dagger}_{{\bf{i}+\hat{\mu}},-\alpha,\downarrow}
-
d^{\dagger}_{{\bf{i}},\alpha,\downarrow}d^{\dagger}_{{\bf{i}+\hat{\mu}},-\alpha,\uparrow})\;, 
\end{split}\end{eqnarray}
or in ${\bf k}$-space,  
$\rm\Delta^{\dagger}({\bf k})$=$\sum_{\alpha} (\cos k_{x}+\cos k_y) 
d^{\dagger}_{{\bf{k}},\alpha,\uparrow}d^{\dagger}_{{\bf{-k}},-\alpha,\downarrow}$. 
$\alpha = x, y$ and ${\hat \mu} = {\hat x}, {\hat y}$.
This operator is a spin singlet that transforms as the $B_{\rm 2g}$ 
representation of the $D_{\rm 4h}$ group \cite{wan},
and it involves different $x$ and $y$ orbitals
on NN sites to optimize the NN kinetic energy [Fig.~4(b)].
In other parts of the phase diagram, a spin-triplet dominates, which
is  odd under orbital exchange, transforms
according to $A_{\rm 2g}$ \cite{wan}, and also involves different orbitals
on NN sites [Fig.~4(c)]. Its projection-1 operator is 
\begin{eqnarray} 
\Delta^{\dagger}({\bf i})_1=\sum_{\hat \mu}(d^{\dagger}_{{\bf i},x,\uparrow} 
d^{\dagger}_{{\bf i}+{\hat \mu},y,\uparrow}- 
d^{\dagger}_{{\bf i},y,\uparrow} 
d^{\dagger}_{{\bf i}+{\hat \mu},x,\uparrow}),
\end{eqnarray}
that in momentum space becomes
$ \rm \Delta^{\dagger}({\bf k})_1$=$(\cos k_x+\cos k_y)
(d^\dagger_{{\bf k},x,\uparrow} d^\dagger_{{\bf -k},y,\uparrow}-
 d^\dagger_{{\bf k},y,\uparrow} d^\dagger_{{\bf -k},x,\uparrow})$.
It resembles the operator of Ref.~\cite{FCZhang}, although they use
on-site pairing. 
Of the
16 possible pairing operators allowed by the symmetry of 
the Hamiltonian \cite{wan,wang}, our singlet and triplet operators
correspond to {\it \#9} and {\it \#12} of Ref.~\cite{wan,future}. 

\begin{figure}
\centerline{
\includegraphics[clip,width=6.5cm]{\mypath{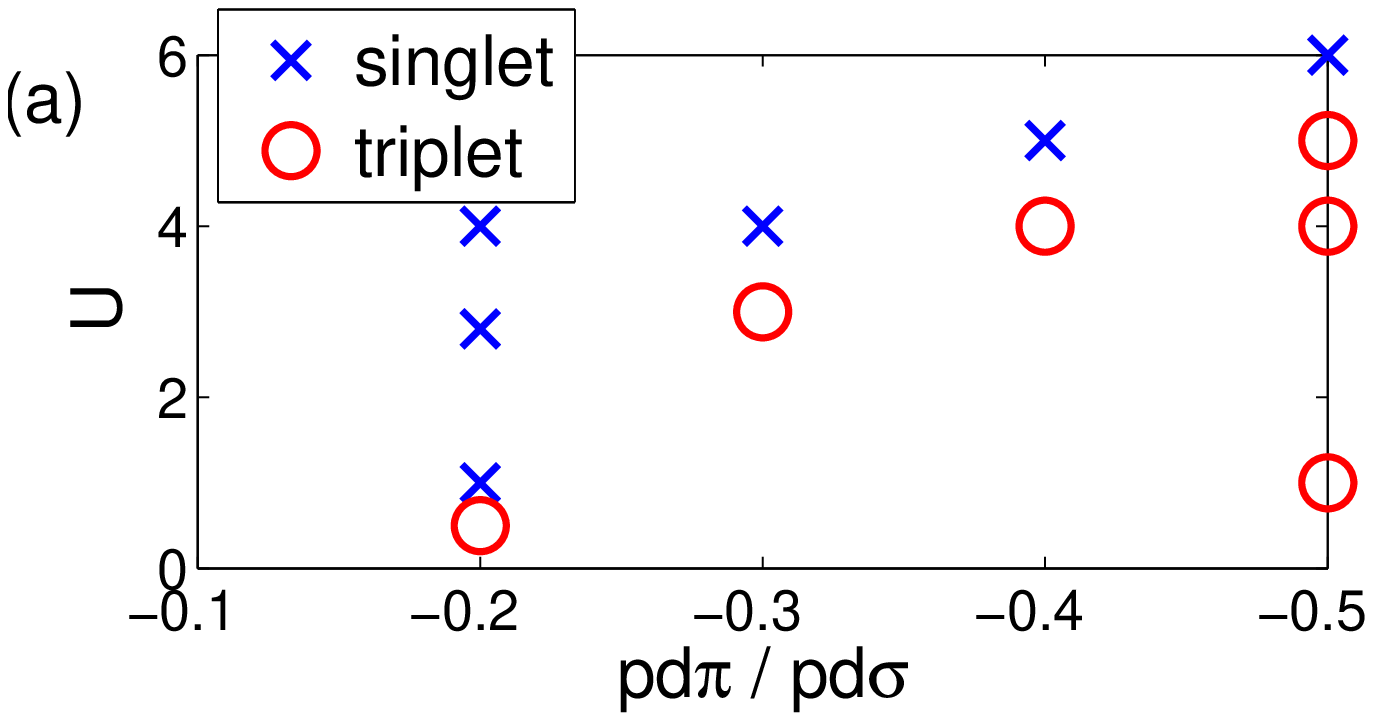}}}
\centerline{
\includegraphics[clip,width=6.0cm]{\mypath{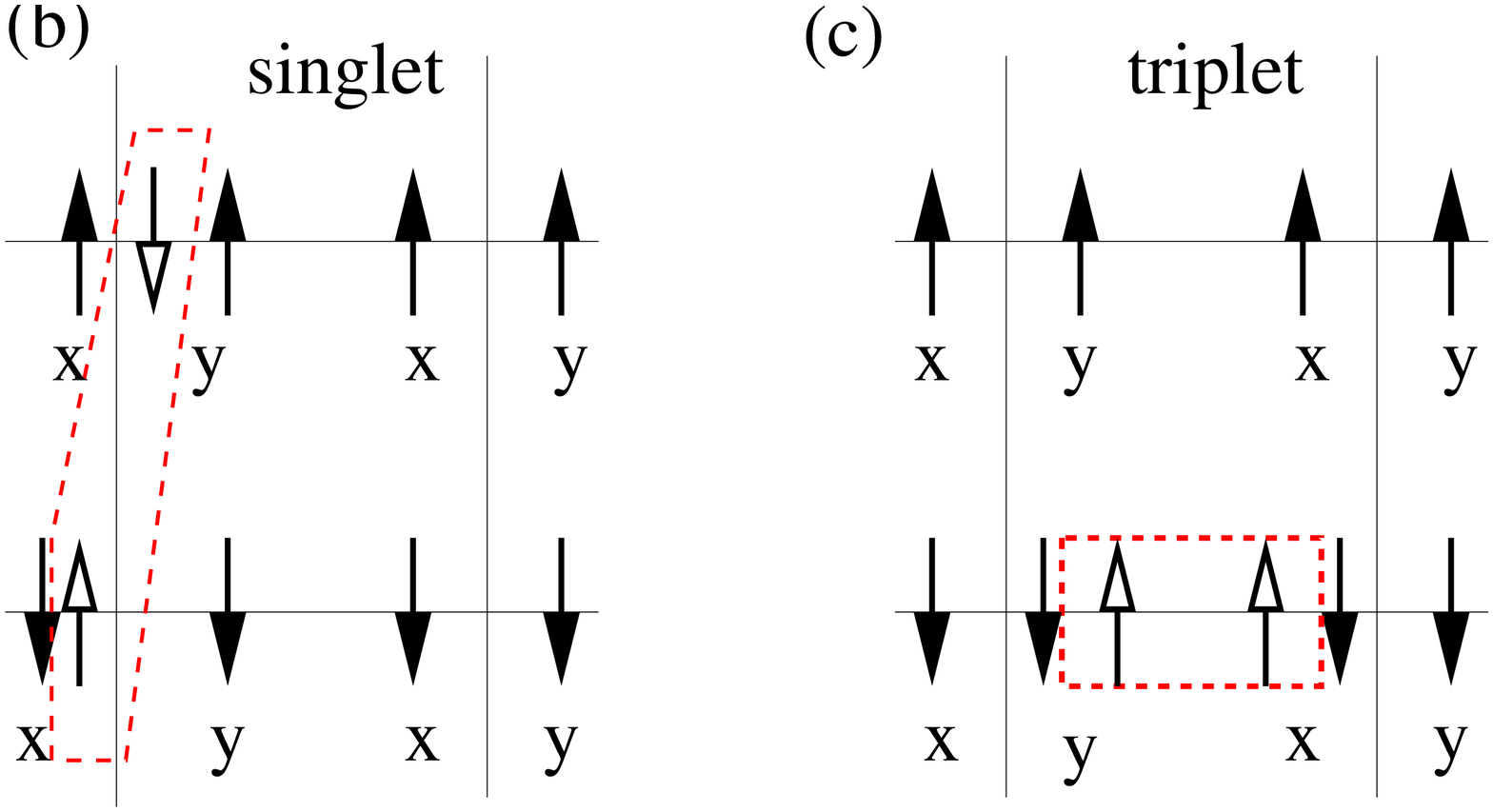}}}
\caption{(Color online) (a) $\sqrt{8}\times\sqrt{8}$ 
cluster results
with 2 more electrons than half-filling, at $U/J$=4. Shown are regions with singlet and triplet
pairing (see text). 
(b,c) Schematic representation of the singlet and triplet pairs (see text). Black arrows represent
the magnetically ordered background. White arrows are the added electrons. 
x,y are the orbitals.
}
\label{Figure4}
\end{figure}

{\it Conclusions:}  
We studied a simple model for the FeAs
superconductors numerically. The undoped system shows $q$$\sim$$(0,\pi)$,$(\pi,0)$ spin order.
We identified dominant pairing operators for two added electrons: depending on parameters they
can be spin singlet or triplet, transforming non-trivially under
$\pi/2$ rotations. Future work will address more realistic models beyond 
two orbitals and contrast their results against those reported here. 



{\it Acknowledgment:}
We acknowledge discussions with F. Reboredo.
Research supported by the NSF grant DMR-0706020, 
the Div. of Materials Sciences and Eng., U.S. DOE under
contract with UT-Batelle, LLC, and the Austrian Science Fund grant P18551-N16.


\begin{thebibliography}{00}
\bibitem{VCAcomment} VCA is run on large but finite 48$\times$48
  lattices. We compared the
  $(0,\pi)$ phase to $(\pi,\pi)$ and ferromagnetic order.
\bibitem{bound} As in ``AF-broken-links'' argumentations for HTSC \cite{RMP},
here pairing of carriers may emerge from the ``damage'' an added electron does to the magnetic background.
\bibitem{future} In a future
publication, we will show that similar results are also obtained using
the parameters of Ref.~\cite{scalapino}.
\end{thebibliography}
\end{document}